\documentstyle[preprint,aps]{revtex}

\begin{document}
%\twocolumn[\hsize\textwidth\columnwidth\hsize\csname @twocolumnfalse\endcsname

\title{\bf Phase transition from a $d_{x^2-y^2}$ to   $d_{x^2-y^2}
+id_{xy}$  superconductor\thanks{To Appear in Physica C}}

\author{Angsula Ghosh and Sadhan K Adhikari\footnote{Corresponding Author
\\ Address: Instituto de F\'{\i}sica Te\'orica, 
Rua Pamplona 145,
01.405-900 S\~ao Paulo, SP, Brazil
\\ Fax: + 55 11 288 8224; 
e-mail: ADHIKARI@IFT.UNESP.BR}}
\address{Instituto de F\'{\i}sica Te\'orica, 
Universidade Estadual Paulista,\\
01.405-900 S\~ao Paulo, S\~ao Paulo, Brazil\\}

\date{\today}
\maketitle

\begin{abstract}

The temperature dependencies of specific heat and  spin susceptibility of a
coupled $d_{x^2-y^2} +id_{xy}$ superconductor in the presence of a weak
$d_{xy}$ component are investigated in the tight-binding model (1) on square
lattice and (2) on a lattice with orthorhombic distortion.  As the
temperature is lowered past the critical temperature $T_c$, first a less
ordered $d_{x^2-y^2}$ superconductor is created, which changes to a more
ordered  $d_{x^2-y^2} +id_{xy}$ superconductor  at $T_{c1} (<T_c)$. 
This manifests
in two second order phase transitions identified by two jumps in specific
heat at $T_c$ and $T_{c1}$.  The temperature  dependencies of the
superconducting observables exhibit a change from power-law to exponential
behavior as temperature is lowered below $T_{c1}$ and confirm the new phase
transition.

{PACS number(s): 74.20.Fg, 74.62.-c, 74.25.Bt}

\end{abstract} 

%\vskip1.5pc]

\newpage

The unconventional high-$T_c$ superconductors \cite{hi} with a high critical
temperature $T_c$ have  a complicated lattice structure with extended and/or
mixed symmetry for the order parameter \cite{n1,n2}.  For many of these
high-$T_c$ materials, the order parameter exhibits anisotropic behavior.  The
detailed nature of anisotropy was thought to 
 be typical to that of an extended
$s$-wave, a pure $d$-wave, or a mixed $(s+\exp(i\theta)d)$-wave type.  Some
high-$T_c$ materials have singlet $d$-wave Cooper pairs  and the order
parameter has $d_{x^2-y^2}$ symmetry in two dimensions \cite{n1,n2}, which has
been supported by recent studies of  temperature dependence of some
superconducting observables
\cite{h1,h2,h3,t1a,t1b,t1c,t1d,c1,c2,c3}. In some cases there is  the
signature  of an extended $s$- or $d$-wave symmetry. The possibility of a
mixed angular-momentum-state symmetry was suggested sometime ago by
Ruckenstein et al.  and Kotliar \cite{6a,6b}. There are experimental evidences
based on Josephson supercurrent for tunneling between Pb and
YBa$_2$Cu$_3$O$_7$ (YBCO)
\cite{5}, and on photoemission studies on Bi$_2$Sr$_2$CaCu$_2$O$_{8+x}$
\cite{7} among others which are difficult to reconcile employing a pure $s$-
or $d$-wave order parameter.  These observations suggest that a mixed
[$s+\exp(i\theta)d$] symmetry is applicable in these cases
\cite{8a,8b}.

More recently Krishana et al. \cite{K} reported a phase transition in the
high-$T_c$ superconductor Bi$_2$Sr$_2$CaCu$_2$O$_8$ induced by a magnetic
field from a study of the thermal conductivity   as a function of temperature
and magnetic field.  Laughlin \cite{L}  have suggested that the new
superconducting phase is the time-reversal symmetry breaking $d_{x^2-y^2}
+id_{xy}$ state. Similar conclusion has been reached by 
Ghosh \cite{G} in a recent work from a study of the temperature dependence of
thermal conductivity.  This has led to the possibility of the
transition to a $d_{x^2-y^2} +id_{xy}$ phase from a pure $d_{x^2-y^2}$ phase
of Bi$_2$Sr$_2$CaCu$_2$O$_8$.  From a study of vortex in a $d$-wave
superconductor using a self-consistent Bogoliubov-de Gennes formalism, Franz
and Te\'sanovi\'c  \cite{F} also predicted the possibility of the creation of a
$d_{x^2-y^2} +id_{xy}$  superconducting state.  Although, the creation of the
mixed superconducting state   $d_{x^2-y^2} +id_{xy}$  is speculative,  Franz
and Te\'sanovi\'c conclude that a dramatic change should be observed (in the
observables of the superconductor) as the superconductor undergoes a phase
transition from a $d_{x^2-y^2}$ state to a $d_{x^2-y^2}+id_{xy}$ state. In
this work we study the effect of this phase transition on superconducting 
specific heat and spin susceptibility in the absence of magnetic field.   The
general trend of observables under the $d_{x^2-y^2} $-  to $d_{x^2-y^2}
+id_{xy}$-wave phase transition of the superconductor, as studied here, is
expected to be independent of the external magnetic field. We recall that
there have been several studies on the formation of a mixed $(s+id)$-wave
superconducting state from a pure $d$-wave 
state \cite{9a1,9a2,9a3,9a4,9a5,9a6}.

First we study the  temperature dependence of the order parameter of the
mixed $d_{x^2-y^2}+id_{xy}$ state within the Bardeen-Cooper-Schrieffer (BCS)
model \cite{e,t}.  The BCS model for a mixed $d_{x^2-y^2}+id_{xy}$ state
becomes  a coupled set of equations.  The ratio of the strengths of the
$d_{x^2-y^2}$- and $d_{xy}$-wave interactions should lie in a narrow region
in order to have a coexisting $d_{x^2-y^2}$- and $d_{xy}$-wave phases in the
case of  $d_{x^2-y^2}+id_{xy}$ symmetry.  As the  $d_{x^2-y^2}$-wave
($d_{xy}$-wave) interaction becomes stronger, the $d_{xy}$-wave
($d_{x^2-y^2}$-wave) component of the order parameter  quickly reduces and
disappears and a pure $d_{x^2-y^2}$-wave ($d_{xy}$-wave) state emerges.

The order parameter of each of $d_{x^2-y^2}$  and $d_{xy}$ states has nodes
on the Fermi surface and may change sign along the Fermi surface. The $s$-wave
order parameter does not have this property. Because of this,
many $d$-wave superconducting observables have power-law dependence
on temperature, whereas the $s$-wave observables exhibit exponential
dependence. We find that in the present coupled $d_{x^2-y^2}+id_{xy}$ state 
 the order parameter does not exhibit nodes and change of sign
along the Fermi surface and exhibits a typical $s$-wave like behavior. 
Consequently, the observables in the coupled $d_{x^2-y^2}+id_{xy}$ state 
does not exhibit typical $d$-wave power-law dependence on temperature,
 but rather a typical $s$-wave exponential dependence.

For a weaker $d_{xy}$-wave admixture,  in the present study we
establish   in the two-dimensional 
tight-binding model (1) on square lattice and (2) on a lattice with 
 orthorhombic distortion
another second-order phase transition at $T=T_{c1}<T_c$, where the
superconducting phase changes from a pure $d_{x^2-y^2}$-wave 
state for $T>T_{c1}$ to a
mixed $d_{x^2-y^2}+id_{xy}$-wave 
 state for $T<T_{c1}$.  The specific heat exhibits two
jumps at the transition points $T=T_{c1}$ and $T=T_c$.  
The temperature
dependencies of the superconducting specific heat and  spin susceptibility
 change drastically at $T=T_{c1}$
from power-law behavior   for $T>T_{c1}$ to exponential behavior
 for $T<T_{c1}$. We find that 
the
observables for the normal state are closer to those for a pure  
superconducting $d_{xy}$
state than to those for a pure
 superconducting $d_{x^2-y^2}$ state.
Consequently, superconductivity in $d_{x^2-y^2}$
 wave is more pronounced than in 
pure $d_{xy}$
wave.  Hence as temperature decreases the system passes from the normal state
to a ``less" superconducting $d_{x^2-y^2}$-wave state  
at $T=T_c$ and then to a ``more"
superconducting  state $d_{x^2-y^2}+id_{xy}$
with dominating $s$-wave behavior
at $T=T_{c1}$ signaling a second
phase transition.

The profound change  in the nature of the superconducting
state at $T=T_{c1}$ becomes  apparent from a study of the entropy. At a 
particular temperature the entropy for the normal state is larger than that
for all superconducting states signaling an increase in order in the
superconducting state. In the case of the present
$d_{x^2-y^2}+id_{xy}$ state we find that
as the temperature is lowered past $T_{c1}$, the entropy of the
superconducting $d_{x^2-y^2}+id_{xy}$
state decreases very rapidly (not shown explicitly
in this work) indicating  the appearance of a more ordered superconducting
phase and a second phase transition.

We base the present study on the two-dimensional  tight binding model 
which we describe below. This model is sufficiently general for considering 
mixed angular momentum states, with or without orthorhombic distortion, 
employing nearest and second-nearest-neighbor hopping integrals.  
The effective interaction in this case can be written as
\begin{eqnarray}\label{2}
V_{{\bf k}{\bf q}}&=& -V_1(\cos k_x-\beta \cos k_y)(\cos q_x
-\beta\cos q_y)\nonumber \\ & - &V_2(\sin k_x\sin k_y)(\sin q_x \sin q_y).
\end{eqnarray}
Here $V_1$ and $V_2$ are the couplings of effective $d_{x^2-y^2}$- and
$d_{xy}$-wave interactions, respectively.  As we shall consider Cooper
pairing and subsequent BCS condensation in both these waves the constants
$V_1$ and $V_2$ will be taken to be  positive  corresponding to attractive
interactions.  In this case the quasiparticle dispersion relation is given by
\begin{equation}\label{3}
\epsilon_{\bf k}=-2t[\cos k_x+\beta \cos k_y-\gamma\cos k_x
\cos k_y],
\end{equation}
where $t$ and $\beta t$ are the nearest-neighbor hopping integrals
along the in-plane $a$ and $b$ axes, respectively, and $\gamma t/2$ is the 
second-nearest-neighbor hopping integral.

We consider the weak-coupling BCS model  in two dimensions with
$d_{x^2-y^2}+id_{xy}$
 symmetry.  At a finite $T$, one has  the following BCS  equation
\begin{eqnarray}
\Delta_{\bf k}& =& -\sum_{\bf q} V_{\bf kq}\frac{\Delta_{ \bf q}}{2E_{\bf
q}}\tanh
\frac{E_{\bf q} }{2k_BT}  \label{130} \end{eqnarray} with $E_{\bf q} =
[(\epsilon_{\bf q} - E_F )^2 + |\Delta_{\bf q}|^{2}]^ {1/2},$ where $E_F$ is
the Fermi energy and $k_B$ the Boltzmann constant. 
The order parameter $\Delta _{\bf q}$ has the
following anisotropic form: $\Delta _{\bf q} \equiv  \Delta_1 (\cos
q_x -\beta \cos q_y) +i\Delta_2 \sin q_x \sin q_y$.    
Using the above form of $\Delta_{\bf q}$ and potential (\ref{2}),
Eq. (\ref{130}) becomes the following coupled set of BCS equations
\begin{eqnarray}
\Delta_1=V_1\sum_{\bf q}\frac{\Delta_1 (\cos q_x-\beta
\cos q_y)^2}{2E_{\bf q}}\tanh
\frac{E_{\bf q}}{2k_BT}\label{131}
\end{eqnarray}
\begin{eqnarray}
\Delta_2=V_2\sum_{\bf q}\frac{\Delta_2(\sin q_x
\sin q_y)^2}
{2E_{\bf q}}\tanh
\frac{E_{\bf q}}{2k_BT}\label{132}
\end{eqnarray}
where the coupling is introduced through $E_{\bf q}$. In Eqs. (\ref{131})
and (\ref{132}) both the interactions $V_1$ and $V_2$ are assumed to be 
energy-independent constants for $|\epsilon_{\bf q} - E_F| < k_B T_D$
and zero for $|\epsilon_{\bf q} - E_F| > k_B T_D$, where $k_B T_D$ is the 
usual Debye cutoff.

The  specific heat   is given by \cite{t}
\begin{equation}
C(T)= -\frac{2}{T}\sum_{\bf q} \frac{\partial f_{\bf q}}{\partial E_{\bf q}}
\left(   E_{\bf q}^2-\frac{1}{2}T \frac{d|\Delta_{\bf q}|^2}{dT} 
 \right)
\label{sp} 
\end{equation} 
where $f_{\bf q}=1/(1+\exp( E_{\bf q}/ k_BT))$.  The spin susceptibility $\chi$
is defined by
\cite{c2,c3}
\begin{equation}
\chi(T)= \frac{2\mu_N^2}{T}\sum_{\bf q}f_{\bf q}(1-f_{\bf q})
\end{equation}
where $\mu_N$ is the  nuclear magneton.

We solved  the coupled set of equations (\ref{131}) and (\ref{132})
numerically by the method of iteration 
and calculated the  gaps $\Delta_1$ and $\Delta_2$ at various 
temperatures for $T<T_c$. We have performed  calculations (1) on a perfect 
square lattice and (2)  in the presence of  an orthorhombic distortion with 
Debye cut off $k_BT_D=0.02586 $ eV ($T_D = 300$ K) in both cases. The 
parameters for these two cases are the following: (1) Square lattice $-$ (a)
$t=0.2586 $ eV, $\beta=1$, $\gamma = 0$, $V_{1}=0.73t$, and $V_2=6.8t$,
$T_c 
= 71$ K, $T_{c1}$ = 28 K; (b) $t=0.2586 $ eV, $\beta=1$,  $\gamma = 0$, 
$V_{1}=0.73t$, and $V_2=7.9t$, $T_c = 71$ K, $T_{c1}$ = 47 K; 
(2) Orthorhombic distortion $-$ (a) $t=0.2586 $ eV, $\beta = 0.95$, and 
$\gamma=0$, $V_{1}=0.97t $, and $V_2=6.5t$, $T_c$ = 70 K, $T_{c1} $ = 25
K;
(b) $t=0.2586 $ eV, $\beta = 0.95$, and $\gamma =0$, $V_{1}=0.97t$,
and $V_2=8.0t$, $T_c$ = 70 K, $T_{c1} $ = 52 K. For a very weak 
$d_{x^2-y^2}$-wave ($d_{xy}$-wave) coupling the only possible solution 
corresponds to $\Delta_2 =0$ ($\Delta_1 =0$).

In Figs.  1  and 2 we plot the temperature dependencies of different
$\Delta$'s for the following two sets of $d_{x^2-y^2}+id_{xy}$-wave
corresponding to models 1  and 2 above (full line $-$ models 1(a) and 2(a);
dashed line $-$ models 1(b) and 2(b)), respectively. In both cases the
temperature dependence of the $\Delta$'s are very similar.  In the coupled
$d_{x^2-y^2}+id_{xy}$-wave as temperature is lowered past $T_c$, the
parameter $\Delta_1$ increases up to $T=T_{c1}.$ With further reduction of
temperature, the parameter $\Delta_2$ becomes nonzero and begins to increase
and eventually both $\Delta_1$ and $\Delta_2$ first increases and then 
saturates as temperature tends
to zero. Recently, the temperature dependencies of the order parameter of the
$d_{x^2-y^2}+is$-wave superconducting state has been studied, where at
$T=T_{c1}$, the transition from $d_{x^2-y^2}$ to $d_{x^2-y^2}+is$ state takes
place. In that case, below $T=T_{c1}$ the $d_{x^2-y^2}$-wave  component of the
order parameter is suppressed, as the $s$-wave component becomes nonzero. No
such suppression of the $d_{x^2-y^2}$-wave takes place in this case as the
$d_{xy}$ component appears.

Now we study the temperature dependence of specific heat in some detail. The
different superconducting and normal specific heats are plotted in Figs. 3
and 4 for square lattice [models 1(a) and 1(b)] and orthorhombic distortion
[models 2(a) and 2(b)], respectively. In both cases the specific heat
exhibits two jumps $-$ one at $T_c$ and another at $T_{c1}$.  From  Eq.
(\ref{sp}) and Figs.  1 and 2 we see that the temperature derivative of
$|\Delta_{\bf q}|^2$ has discontinuities at $T_c$ and $T_{c1}$ due to the
vanishing of $\Delta_1$ and $\Delta_2$, respectively, responsible for the two
jumps in specific heat.  For a pure $d_{x^2-y^2}$ wave we find that the
specific heat exhibits a power-law dependence on temperature. However,  the
exponent of this dependence varies with temperature. For small $T$ the
exponent is approximately 2.5, and for large $T$ ($T\to T_c$) it is nearly 2.
In the $d_{x^2-y^2}+id_{xy}$ model, for $T_c > T > T_{c1}$ the specific heat
exhibits $d_{x^2-y^2}$-wave power-law behavior; for  $ T < T_{c1}$ the
specific heat exhibits an $s$-wave like exponential behavior.  For the
$d$-wave model $d_{x^2-y^2}$, $C_s(T_c)/C_n(T_c)$ is a function of $T_c$ and
$\beta$. In Figs. 3  and 4  this ratio,  for $T_c$ = 70 K, is approximately 3
(2.5) for $\beta =$ 1 (0.95).  In a continuum $d$-wave 
calculation  this ratio was 2
in the absence of a van Hove singularity \cite{c2,c3}. We also calculated the
specific heat for the pure $d_{xy}$ case. For square lattice
 with $V_1 =9.0$ we obtain $T_c =67$ K  and $C_s(T_c)/C_n(T_c)=1.82$
and $C_s(T)/C_n(T_c)
\sim (T/T_c)^{1.4}$ for the whole temperature range.
For orthorhombic distortion $\beta
=0.95$, with $V_1 =9.0$ we obtain $T_c =69$ K  and $C_s(T_c)/C_n(T_c)=1.94$
and $C_s(T)/C_n(T_c)
\sim (T/T_c)^{1.5}$ for the whole temperature range. 
This power-law behavior 
with temperature in both the $d$ waves is destroyed in the coupled 
  $d_{x^2-y^2}+id_{xy}$ wave and 
for $T<T_{c1}$, we find an $s$-wave-like exponential behavior in both
cases.  In both the uncoupled $d$ waves the order parameter $\Delta$ has
nodes on the Fermi surface and changes sign and this property is destroyed in
the coupled  $d_{x^2-y^2}+id_{xy}$  wave, where the order parameter has a
typical $s$ wave behavior.

In Fig. 5 we study the jump $\Delta C$ in the specific heat at $T_c$ for pure
$s$- and $d$-wave superconductors as a function of $T_c$, where we plot
the ratio 
$\Delta C/C_n(T_c)$ versus $T_c$. For a BCS superconductor in the continuum
$\Delta C/C_n(T_c)$ = 1.43  (1.0) for $s$-wave ($d$-wave) 
independent of $T_c$ \cite{c2,c3,t}. 
Because of the presence of the van Hove singularity in
the present model this ratio increases with $T_c$ as can be seen in Fig. 5.
For a fixed $T_c$, the ratio 
$\Delta C/C_n(T_c)$ is larger for square lattice ($\beta =1$) than 
that for a lattice with orthorhombic distortion ($\beta=0.95$) for both $s$ 
and $d_{x^2-y^2}$ waves. However, for a $d_{xy}$-wave superconductor
 $\Delta C/C_n(T_c)$ is smaller for square lattice ($\beta =1$) than 
that for a lattice with orthorhombic distortion ($\beta=0.95$). The jump in
$d_{xy}$ wave is smaller than that for $s$ and $d_{x^2-y^2}$ waves.
At $T_c$ = 100 K,  in the  $s$-wave ($d_{x^2-y^2}$-wave)
square lattice case this ratio could be as high as 3.63 (2.92), whereas for 
$d_{xy}$ wave this ratio at 100 K is 1.15 (1.25) for square lattice
(orthorhombic distortion). 

Next we study the temperature dependencies of spin susceptibility for square
lattice and in the presence of orthorhombic distortion
 which we exhibit in Figs. 6 and 7, respectively.
There  we  plot the results for pure $d_{x^2-y^2}$, $d_{xy}$, and $s$ waves  
for comparison, in addition to those for models 1(a), 1(b), 2(a) and 2(b). 
In all cases reported in these figures $T_c \approx 70$ K. 
For pure $d$-wave cases  we obtain
power-law dependencies on temperature. The exponent for this power-law
scaling is independent of critical temperature $T_c$ but vary from a
square lattice to that with an orthorhombic distortion. 
For $d_{x^2-y^2}$ wave, the exponent for square
lattice  (orthorhombic distortion, $\beta$ = 0.95)   is 2.6 (2.4).  
For $d_{xy}$ wave, the exponent for square
lattice  (orthorhombic distortion, $\beta$ = 0.95)   is 1.1 (1.6).  
For the mixed  $d_{x^2-y^2}+id_{xy}$ wave, 
$d_{x^2-y^2}$-wave power-law behavior is obtained for
$T_c>T>T_{c1}$.  For
$T<T_{c1}$,  one has a typical  $s$-wave behavior.  

In conclusion, we have studied the ($d_{x^2-y^2}+id_{xy}$)-wave 
superconductivity employing
a two-dimensional tight binding BCS model on square lattice and also for 
orthorhombic distortion.  We
have kept the potential  couplings in such a domain that a coupled
($d_{x^2-y^2}+id_{xy}$)-wave 
solution is allowed. For a weaker $d_{xy}$ admixture,
as temperature is lowered past the  first
critical temperature $T_c$, a weaker (less ordered) superconducting phase is
created  in $d_{x^2-y^2}$ wave, which changes to a stronger (more ordered)
superconducting phase in ($d_{x^2-y^2}+id_{xy}$) wave at $T_{c1}$. 
    The $d_{x^2-y^2}+id_{xy}$-wave
state is similar to an   $s$-wave-type  state with no node in the order
parameter. The phase transition at $T_{c1}$ from a $d_{x^2-y^2}$ wave
to a $d_{x^2-y^2}+id_{xy}$ wave
is  marked by power-law
(exponential) temperature dependencies of specific heat and spin susceptibility
 for $T > T_{c1}$ ($<T_{c1} $).  Similar behavior has been observed for a
 $d_{x^2-y^2}+is$-wave state \cite{9a1}. 
 As the mixed state is $s$-wave like in both 
 cases, from the present study it would not be possible to identify the
proper symmetry of the order parameter $-$ $d_{x^2-y^2}+id_{xy}$
opposed to $d_{x^2-y^2}+is$ $-$ and further phase sensitive tests of
pairing symmetry in cuprate superconductors is needed.

We thank Conselho
Nacional de Desenvolvimento Cient\'{\i}fico e Tecnol\'ogico and Funda\c c\~ao
de Amparo \`a Pesquisa do Estado de S\~ao Paulo for financial support.

\newpage

{\bf Figure Captions:}
\vskip 1cm

1. The  order parameters
$\Delta_1$,  $\Delta_2$ in Kelvin 
 at different temperatures for $d_{x^2-y^2}+id_{xy}$-wave models 
1(a) (full line) and 1(b)  (dashed line) for square lattice.

2. The  order parameters
$\Delta_1$,  $\Delta_2$ in Kelvin 
 at different temperatures for $d_{x^2-y^2}+id_{xy}$-wave models 
 2(a) (full line)  and 2(b) (dashed line)
  in presence of orthorhombic distortion
$(\beta = 0.95)$.

3. Specific heat ratio $C(T)/C_n(T_c)$ versus $T/T_c$  for models 1(a)
and 1(b) for square lattice:
1(a) (full line), 1(b)
 (dashed line), $d_{xy}$ (dotted line), normal (dashed-dotted line).
In all cases $T_c \approx 70$ K.

 4. Specific heat ratio $C(T)/C_n(T_c)$ versus $T/T_c$  for models 2(a)
and 2(b) for orthorhombic distortion:
2(a) (full line), 2(b)
 (dashed line), $d_{xy}$ (dotted line), normal (dashed-dotted line).
In all cases $T_c \approx 70$ K.

5. Specific heat jump for different $T_c$ for pure $s$ and $d$ waves:
$s$ wave (solid line, square lattice),
$s$ wave (dashed line, orthorhombic distortion),
$d_{x^2-y^2}$ wave (dashed-dotted line, square lattice),
$d_{x^2-y^2}$ wave (dashed-double-dotted line, orthorhombic distortion),
$d_{xy}$ wave (dotted line, square lattice), $d_{xy}$ wave
(dashed-triple-dotted line, orthorhombic distortion).
 
 6. Susceptibility  ratio $\chi(T)/\chi(T_c)$ for square lattice
 versus $T/T_c$:  
 pure $d_{x^2-y^2}$ wave (solid line), pure $d_{xy}$ wave (dashed line),
pure $s$ wave (dashed-dotted line),
 model 1(a) (dotted line),
 model 1(b) (dashed-double-dotted line).
In all cases $T_c \approx 70$ K. 

7. Susceptibility  ratio $\chi(T)/\chi(T_c)$ in presence of orthorhombic
distortion 
 versus $T/T_c$:  
  pure $d_{x^2-y^2}$ wave (solid line), pure $d_{xy}$ wave (dashed line),
pure $s$ wave (dashed-dotted line),
 model 2(a) (dotted line),
 model 2(b) (dashed-double-dotted line).
In all cases $T_c \approx 70$ K.

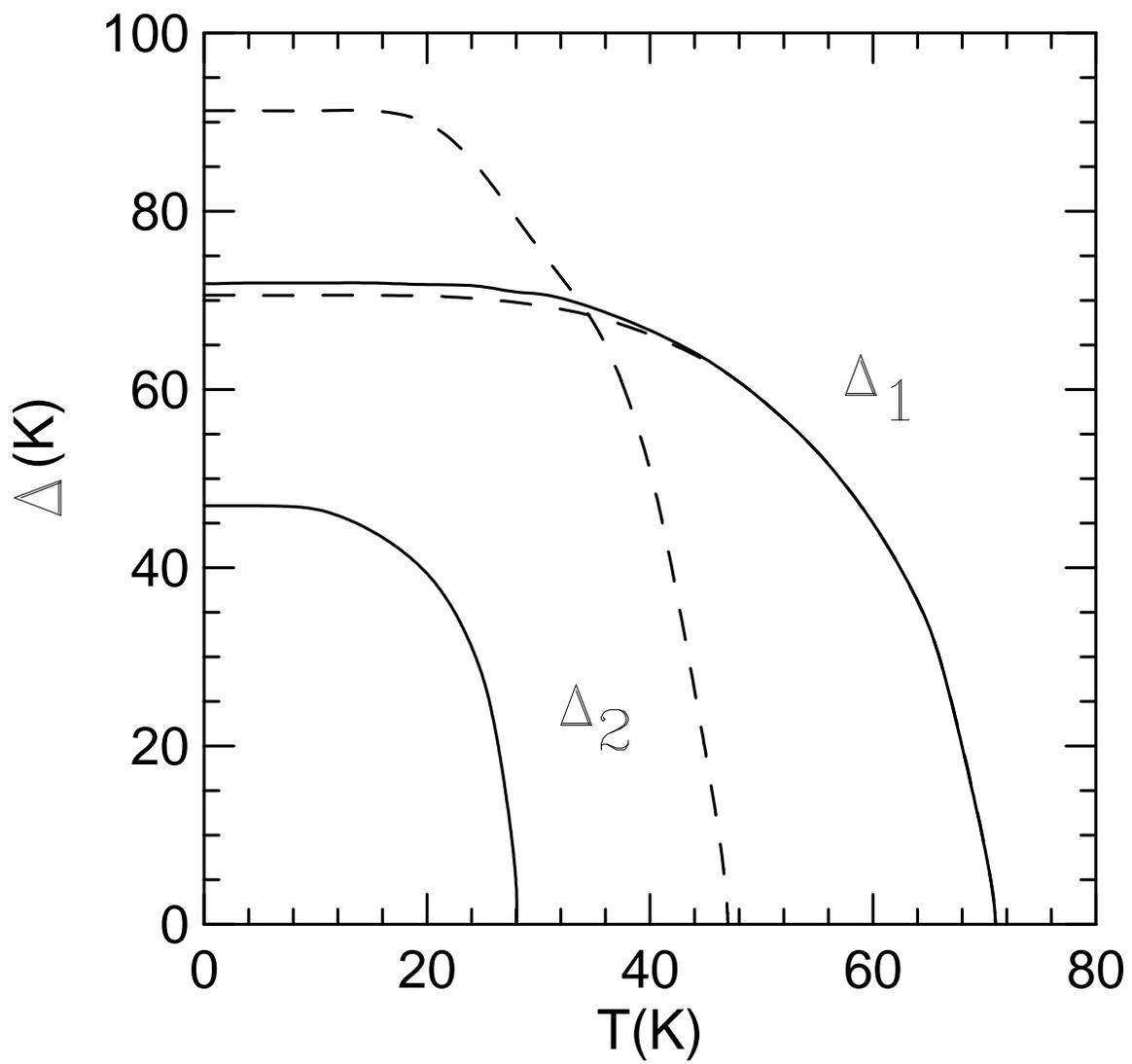
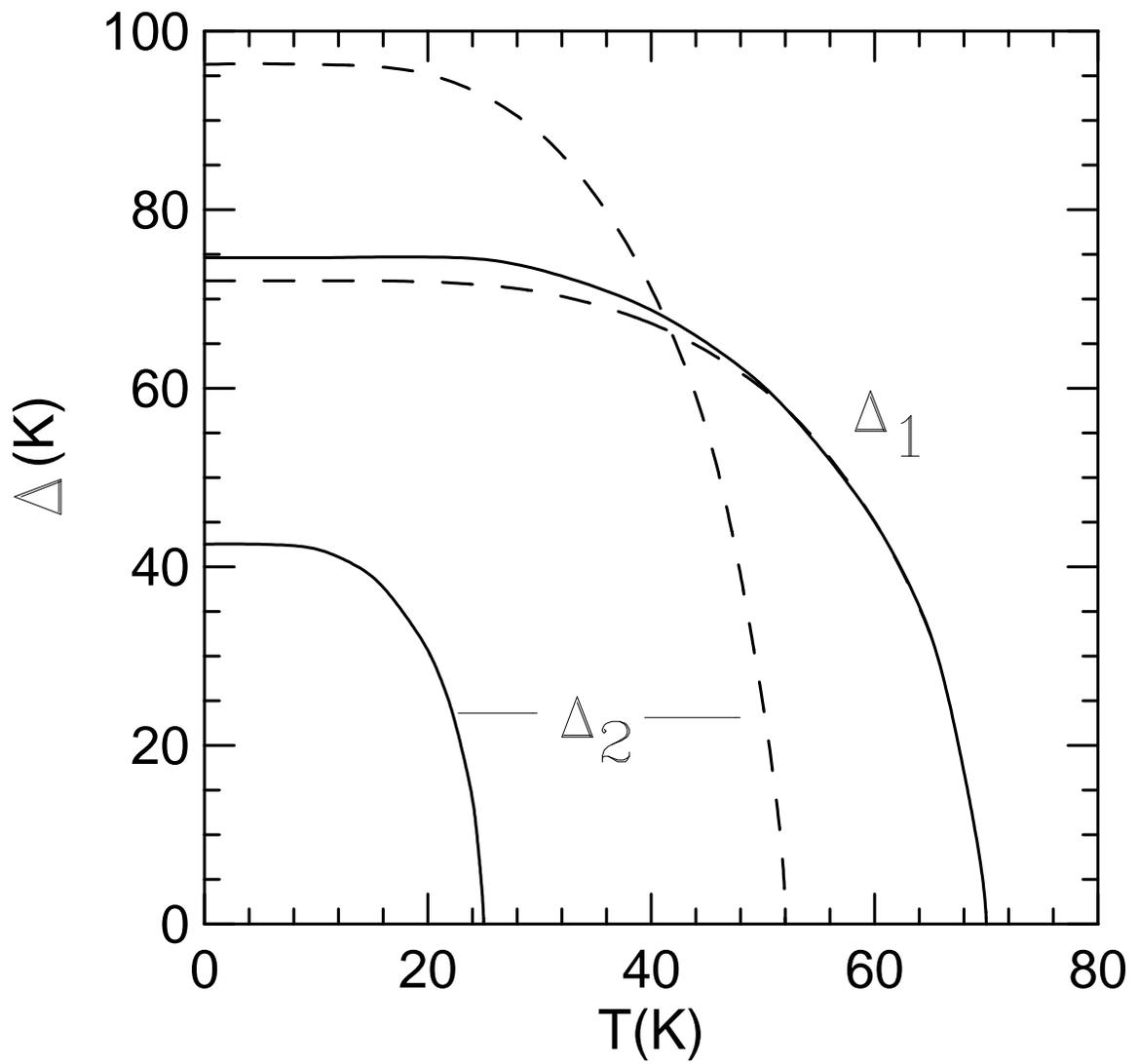
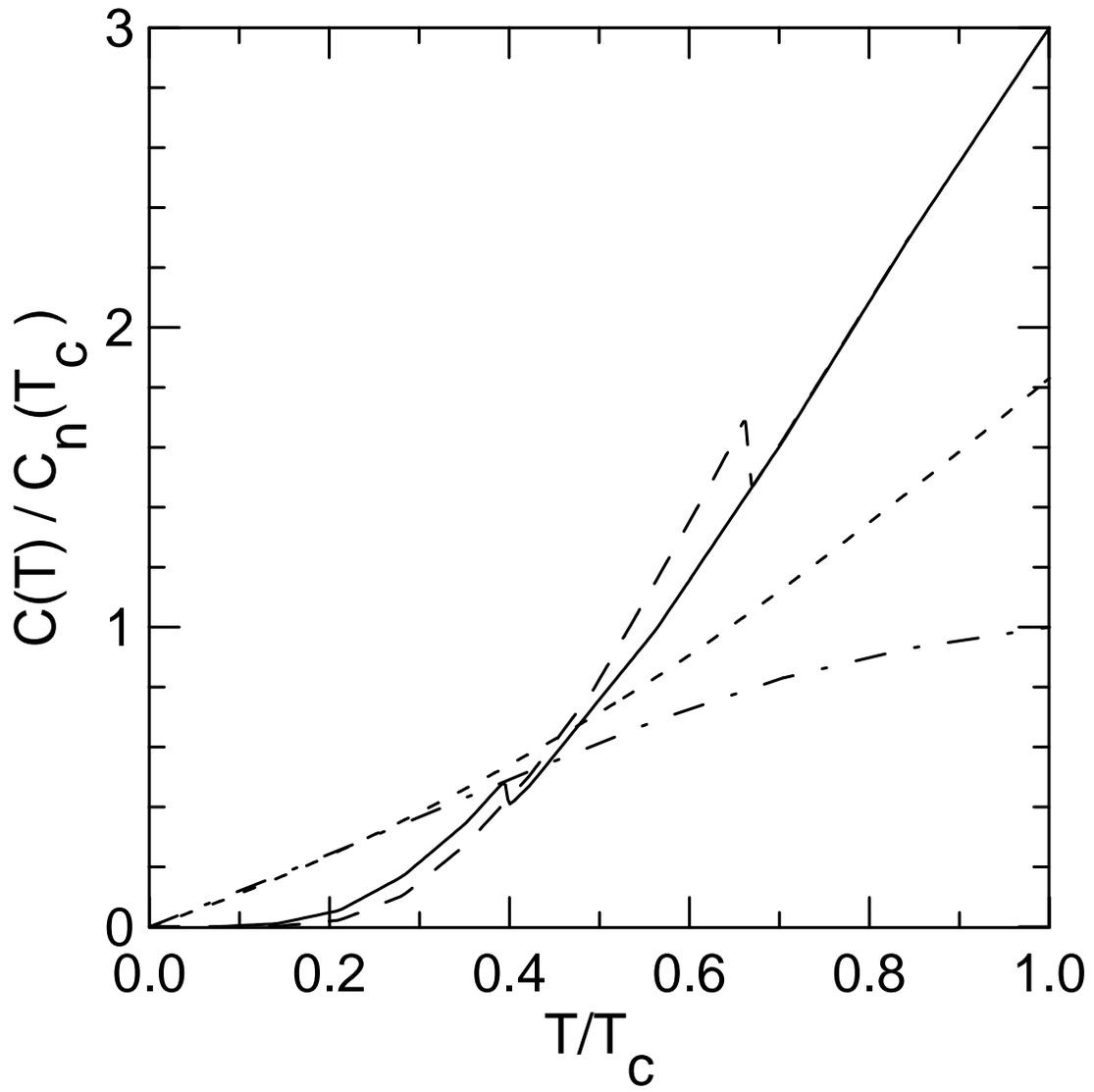
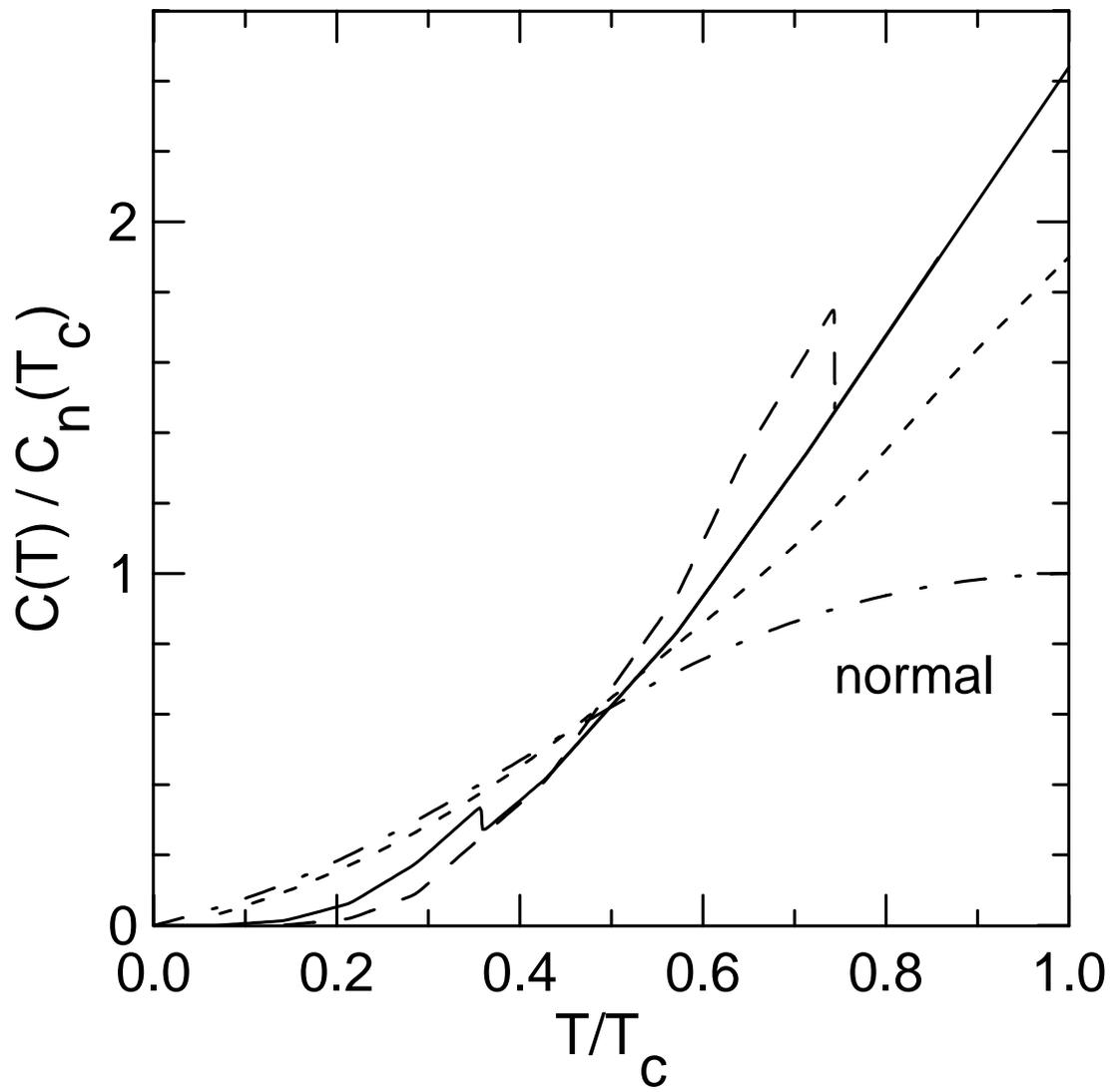
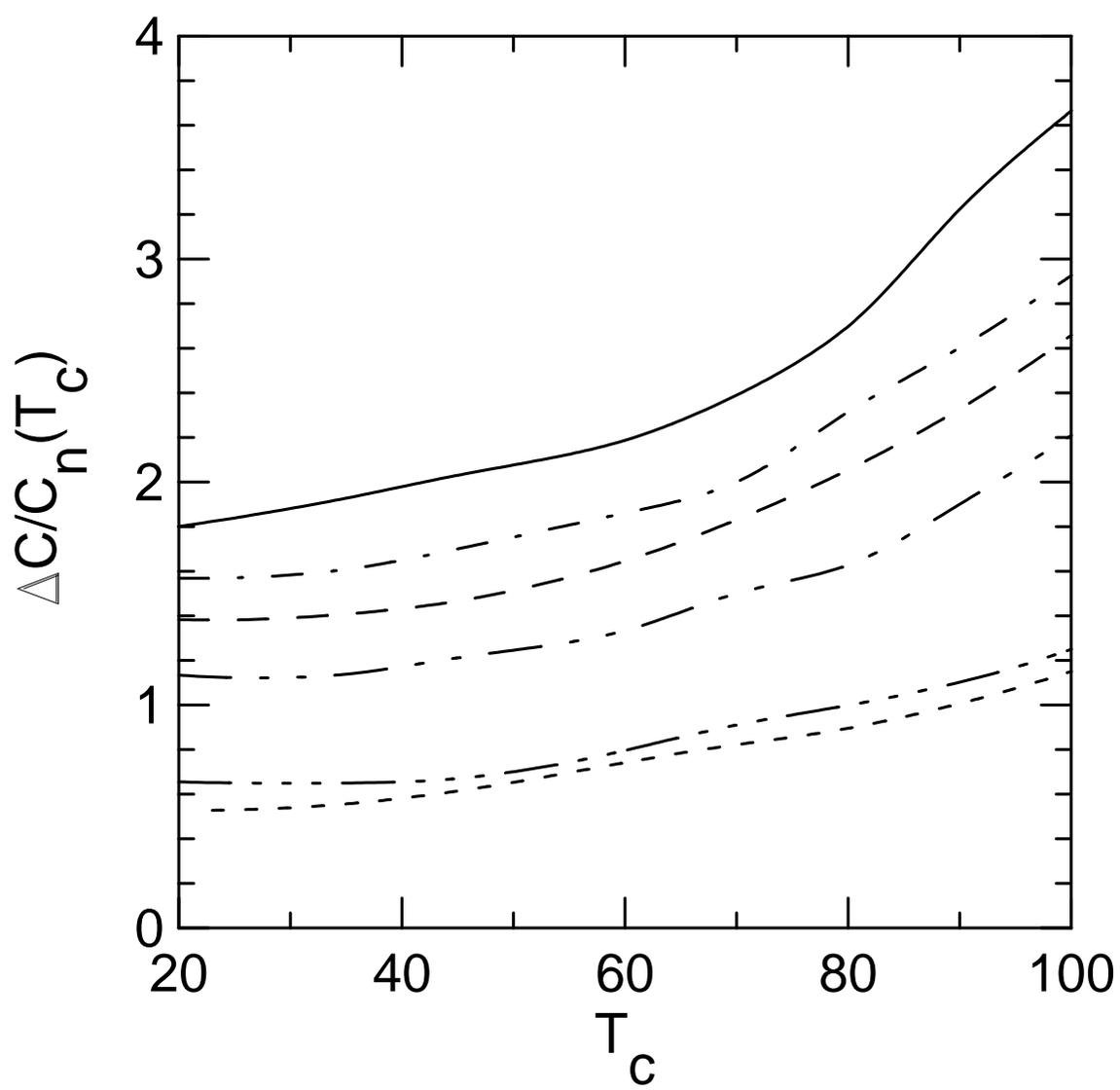
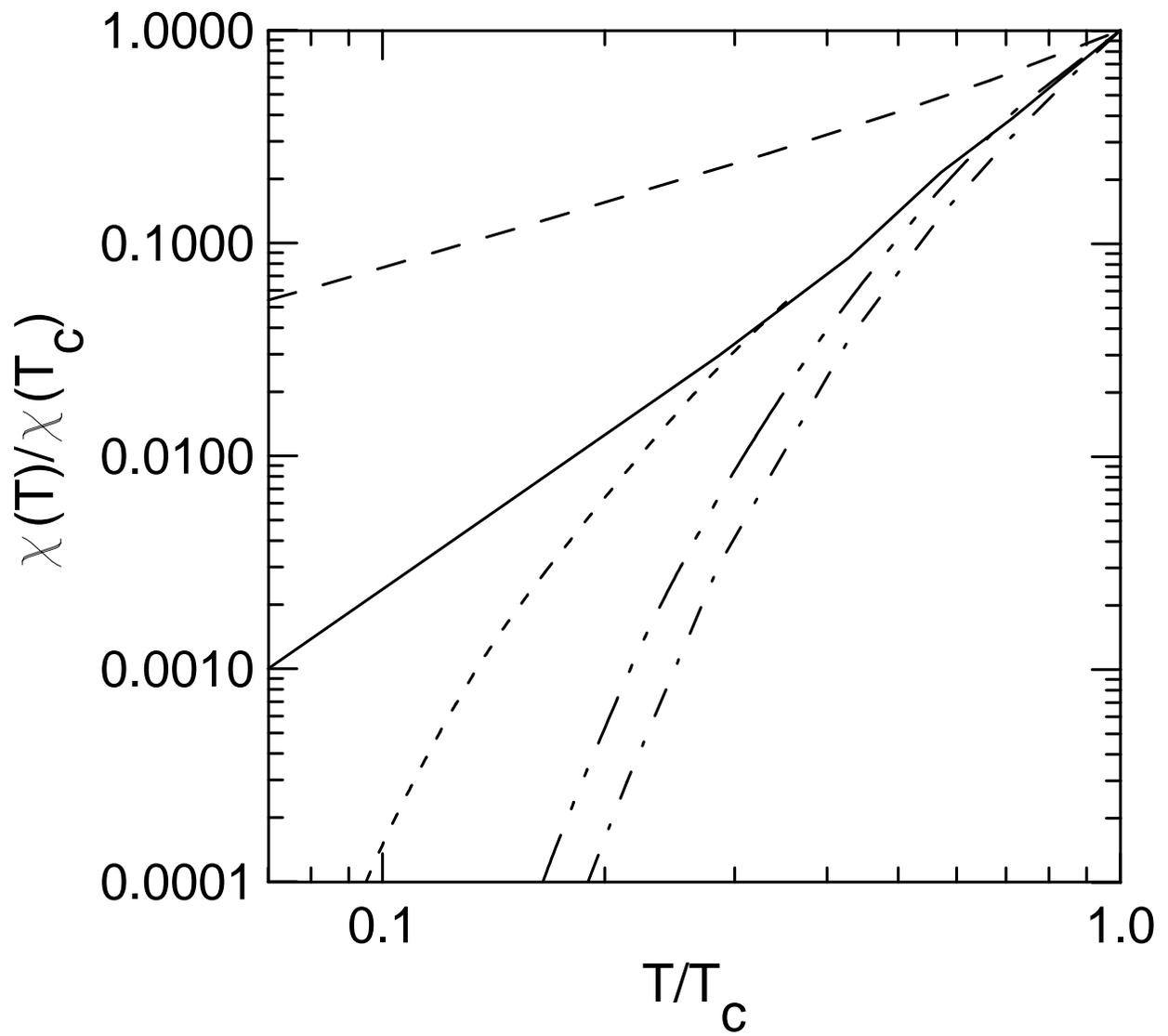
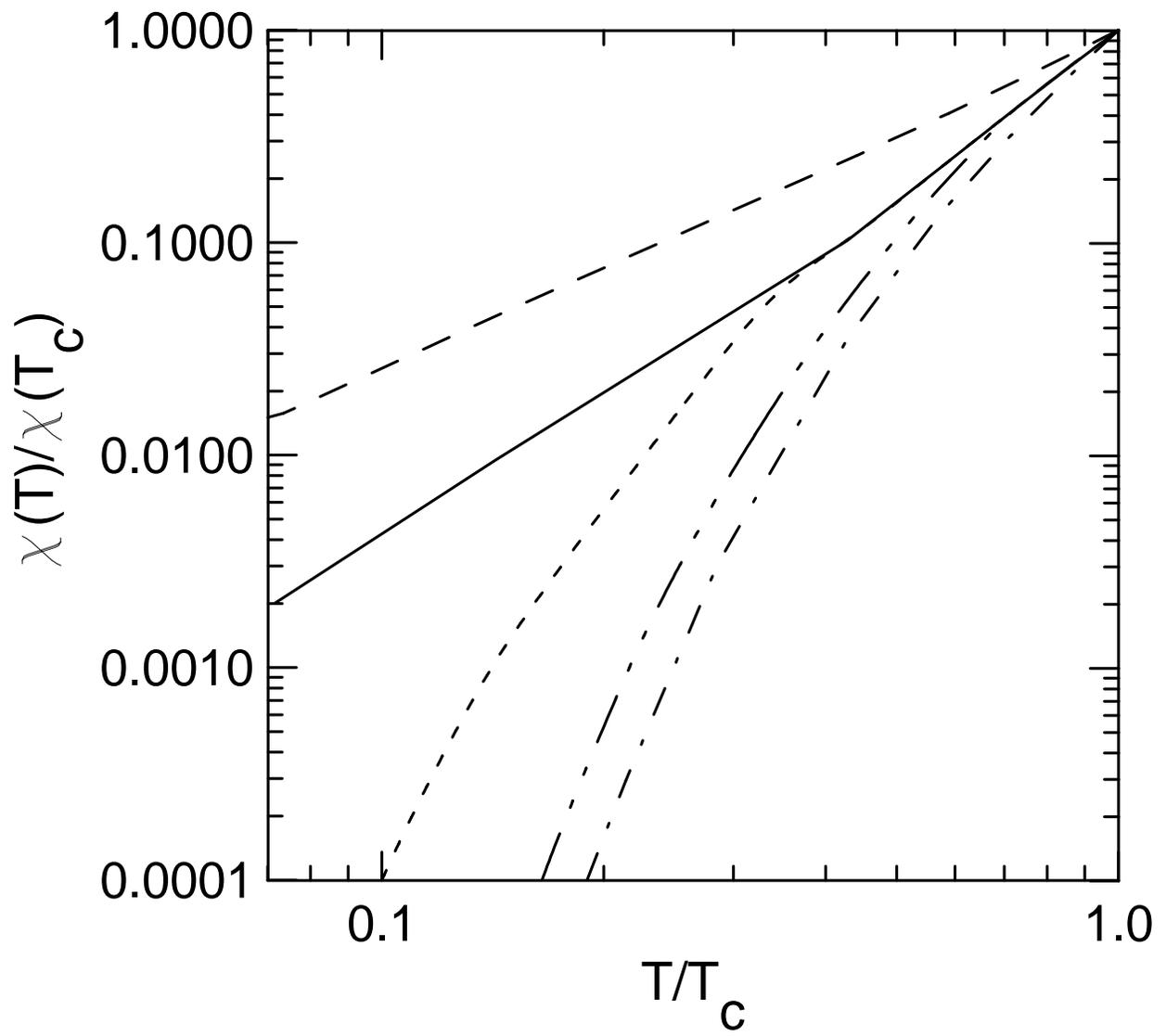
\end{document}